\documentclass[a4paper]{article}

\usepackage{INTERSPEECH2021}
\usepackage{amsmath,amssymb,amsfonts}
\usepackage{bm}
\usepackage{algorithmic}
\usepackage{graphicx}
\usepackage{textcomp}
\usepackage{xcolor}
\usepackage{booktabs}
\usepackage[hidelinks]{hyperref}
\title{LEAP Submission for the Third DIHARD Diarization Challenge}
\name{Prachi Singh, Rajat Varma, Venkat Krishnamohan, Srikanth Raj Chetupalli, Sriram Ganapathy \thanks{This work was supported by the grants from  the British Telecom Research Center.}}

\address{Learning and Extraction of Acoustic Patterns (LEAP) Lab,\\
	Electrical Engineering, Indian Institute of Science, Bangalore, India.
}
\email{\{prachisingh, rajatvarma, venkatk, sraj, sriramg\}@iisc.ac.in}

\begin{document}
	
	\maketitle
	\begin{abstract}
		The LEAP submission for DIHARD-III challenge is described
		in this paper. The proposed system is composed of a speech bandwidth classifier, and diarization systems fine-tuned for narrowband and wideband speech separately. We use an end-to-end speaker diarization system for the narrowband conversational telephone speech recordings. For the wideband multi-speaker recordings, we use a neural embedding based clustering approach, similar to the baseline system. The embeddings are extracted from a time-delay neural network (called x-vectors) followed by the graph based path integral clustering (PIC) approach. The LEAP system showed 24\% and 18\%
		relative improvements for Track-1 and Track-2 respectively over
		the baseline system provided by the organizers. This paper describes the challenge submission, the post-evaluation analysis and improvements observed on the DIHARD-III dataset.
		% report provides details of the system and the experimental results on the DIHARD-III dataset.
	\end{abstract}
	
	\section{Introduction}
	The task of identifying ``who spoke when'' in a speech recording, referred to as ``Speaker diarization'', is important for speech analytics applications such as rich transcription, content retrieval and indexing, etc. Speaker diarization is challenging due to the source related factors, such as the number of speakers in the recording, nature of the conversation, turn taking behavior, and the channel related factors such as the conversation medium, presence of noise and reverberation etc. The DIHARD series of challenges \cite{ryant2018first,ryant2019second,ryant2020dihard} are designed to benchmark the performance of speaker diarization systems on a common set of speech recordings collected from diverse domains. In the current paper, we describe the LEAP lab submission to the DIHARD-III challenge \cite{ryant2020dihard}. 
	\par The popular approach, including the baseline system for the DIHARD-III challenge \cite{ryant2020dihard, singh2019leap}, comprises of (i) speech activity detection and segmentation of audio, (ii) extraction of speaker embeddings, (iii) pair-wise similarity scoring, (iv) clustering, and (v) resegmentation steps to generate the diarization output for a given recording. An example state-of-the-art system \cite{ryant2020dihard} uses embeddings extracted from a neural network (x-vectors) \cite{snyder2019speaker}, probabilistic linear discriminant analysis (PLDA) model \cite{4409052} for scoring \cite{silovsky2011plda}, agglomerative hierarchical clustering (AHC) \cite{Jordi2012}, and a variational-Bayes hidden Markov model (VB-HMM) for resegmentation \cite{diez2018speaker}. However, a single speaker is assumed in each segment, which limits the performance of the systems designed in this approach on recordings with overlap. In other approaches \cite{Bullock2020}, a separate overlap detection system is employed to assign more than one speaker label to the overlapping regions during resegmentation.
	
	The end-to-end neural diarization (EEND) systems handle the speaker overlap naturally, and have gained interest in the recent past \cite{shinji2019ASRU, shinji2020Interspeech}. Short-term features with context and frame sub-sampling are input to a neural network, trained with a permutation invariant binary cross-entropy objective for each speaker activity. The presence of unknown number of speakers is handled using encoder-decoder based attractor \cite{shinji2020Interspeech} modeling. 
	% However, training the EEND system with more speakers is challenging due to the number of permutations involved in the loss computation. 
	% Due to the variability in the channel and source related factors, a single modeling approach may not work, and hence a combination of several systems is used in practice. The system combination can be based on the features, embeddings, similarity score or even of the output Rich Transcription Time Marked (RTTM) file. Recently, RTTM level combination \cite{Raj2021Doverlap} has gained interest and shown to be of great help in improving the diarization system performance. 
	
	In the LEAP submission, we explore different systems for the narrowband and wideband speech recordings, overlap processing using an off-the-shelf overlap detector.
	% , and system combination at the score level. 
	We have trained an end-to-end system for the telephone recordings in narrowband.  As number of speakers varies from 1-10 in the wideband recordings in DIHARD, end-to-end training becomes computationally tedious and time consuming due to the number of permutations involved in the loss computation. Therefore, we have explored alternative graph based clustering approach known as path integral clustering (PIC) for diarization inspired by \cite{singh2021pic}. 

	\section{Detailed description of the systems}\label{sec:description}
	
\begin{figure}[t!]
		\centering
		\includegraphics[trim={4cm 0cm 10cm 0cm},clip,width=\linewidth]{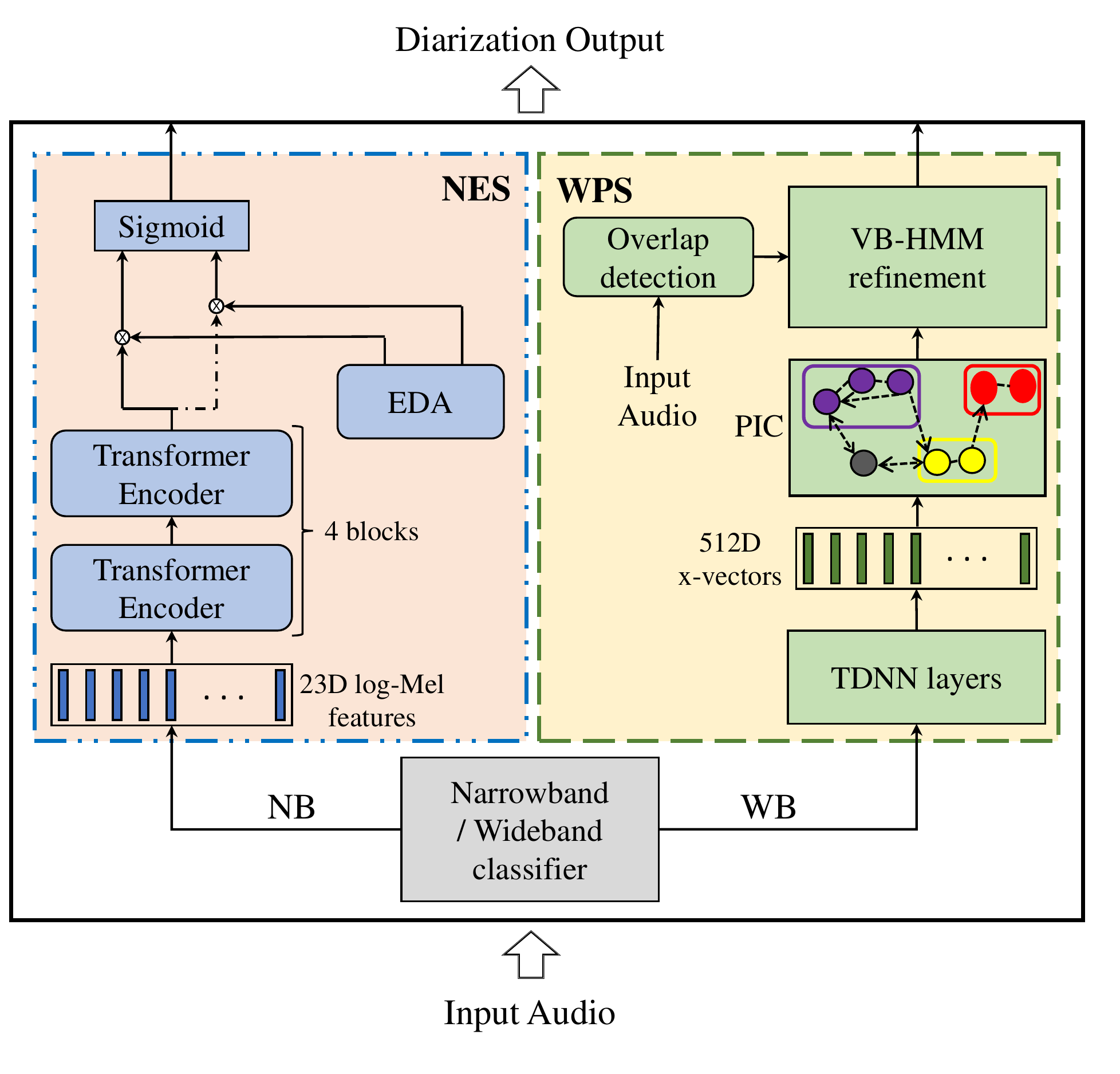}
		\caption{Overall Block Diagram of LEAP system submitted to DIHARD III.}
		\label{fig:block_diagram}
	\end{figure}

	% \subsection{Baseline system}\label{sec:baseline}
	% The baseline systems for both the tracks are implemented as described in \cite{ryant2020dihard}. We have used the best system configuration obtained for Track-1 along with the pre-trained baseline SAD model for Track-2.
	The overall block schematic of the LEAP system is given in Figure \ref{fig:block_diagram}. The input audio is passed through a narrowband/wideband classifier. If the classifier output is narrowband (NB), the audio is passed through the narrowband end-to-end system (NES), otherwise a wideband PIC system (WPS) is used for diarization. The details of each independent system are discussed in the following sections.
	
	\subsection{Wideband-Narrowband classifier}
	A two-layer fully-connected feed-forward neural network with x-vectors as input features is used as the wideband-narrowband classifier. X-vectors are extracted using baseline 
	TDNN model \cite{ryant2020dihard} containing 5 TDNN layers followed by statistics pooling layer which is followed by 2 fully connected layers as described in \cite{snyder2019speaker}. The model is trained using Voxceleb1 and 2 datasets .The input x-vectors are $512$-dimensional and extracted every $5$s using segments of duration $10$s. During inference, a majority voting of the classifier neural network's prediction for all the segments is used to make a file-level decision on the recording bandwidth. 
	
	\subsection{Wideband PIC system (WPS)}\label{sec:wps}
	The diarization system for wideband speech recordings is inspired by the multi-stage baseline \cite{ryant2020dihard}. The WPS consists of neural embedding extraction followed by pair-wise similarity scoring and clustering of short speech segments and VB-HMM re-segmentation with overlap processing. In our setup, we have explored two models for embedding extraction.
	
	\subsubsection{Embedding extraction}\label{sec:ee}
	We use x-vectors as the embeddings, which are extracted using two architecture variants, (i) extended-TDNN (ETDNN) and (ii) factorized-TDNN (FTDNN). We used 30-dimensional mel frequency cepstral coefficient (MFCC) features, extracted every $10$~ms using a $25$~ms window, as inputs to both the models. Both models are trained on the VoxCeleb1 \cite{nagrani2017voxceleb} and VoxCeleb2 \cite{Chung2018} datasets using speaker identification as the objective to discriminate among $7,320$ speakers. We extract the embeddings using a segment size of $1.5$~s and a temporal shift of $0.25$~s for both the models.
	
	\textbf{ETDNN}: The 14-layer ETDNN model follows the architecture described in \cite{snyder2019speaker}. The model has 4 TDNN layers with 1 fully connected layer and ReLu activation after each TDNN layer of $512D$, with a total temporal context of $\pm 11$ neighboring frames. Following the TDNN layers, the model has two feed-forward layers containing $\{512,1500\}$ units followed by mean and standard deviation pooling layer of 3000D. The $512$ dimensional output of the affine component of the $12^{th}$ layer before non-linearity is taken as the x-vector embedding. 
	
	\textbf{FTDNN}: The architecture of the 14-layer FTDNN model is similar to that of ETDNN, with factorized TDNN layers \cite{Povey2018} in place of the TDNN layers. The model has a reduced number of parameters due to the factorization of each layer's weight matrix $M$ as a product of two low-rank matrices. 
	% obtained using singular value decomposition (SVD). 
	The model enforces second factor in $M=AB$ to be semi-orthogonal ($MM^T=I$). The factorized TDNN layers are each of size 1024, and perform a sequence of convolutions that maps the input to 256 dimensions and back to the hidden layer size of 1024. The network has a longer temporal context of $\pm 16$ frames. We extract $512$-dimensional output from the $12^{th}$ affine layer of the model as the x-vector embedding
	
	\subsubsection{Similarity Scoring}
	We consider (i) cosine score and (ii) PLDA score to compute the similarity between segment level x-vector features. To compute the cosine score, the x-vectors are first projected into a $30$-dimensional space using principle component analysis (PCA), computed using the development dataset and the score is computed in the projected space. The probabilistic linear discriminant analysis (PLDA) models are trained separately for the ETDNN and FTDNN x-vectors, and a binary hypothesis testing framework is used to obtain the PLDA similarity score between a pair of segments. The x-vectors extracted from subset of Voxceleb1 and Voxceleb2 containing 128k utterances and 7,200 speakers are used for training the PLDA model. We apply a recording level PCA on the x-vectors reducing the dimension based on 30\% of total energy.
	
	\subsubsection{Path integral clustering}\label{sec:pic}
	We explored a graph-structural-based agglomerative clustering algorithm, known as path integral clustering (PIC) \cite{zhang2013agglomerative}, using the segment similarity scores to obtain the diarization output. The clustering process involves the creation of a directed graph $G=(V, E)$, where input features are the vertices $(V)$ and $E$ is the set of edges connecting the vertices.  PIC merges clusters in a bottom-up manner using affinities computed using the across cluster edge/path weights defined as incremental path integral. Higher affinity indicates dense connections across clusters. 
	% Here each cluster is treated as dynamical system and path integral is the structural descriptor of cluster.
	
	Let $\bm{X}=\{x_1,x_2,...,x_{N_r}\}$ be the set of $N_r$ x-vector embeddings extracted from a recording $r$. We first compute a sparse adjacency matrix $\bm{W}\in\mathbb{R}^{N_r\times N_r}$ of graph $G$ based on the $K$-nearest neighbours of each $x_i$. We use sigmoid of similarity scores (cosine/PLDA) as edge weights in $\bm{W}$. 
	% The transition probability matrix $\boldsymbol{P}$ is obtained by normalizing each row of $\bm{W}$ by its sum. 
	For initialization, embeddings are clustered based on their nearest neighbour.
	% Similar to the agglomerative hierarchical clustering (AHC), PIC also merges two clusters at each time step based on maximum affinity, which is computed using the path integral as defined in \cite{zhang2013agglomerative}. 
	The cluster affinity measure between two clusters for the PIC algorithm is computed as,
	\begin{equation}\label{eq:affnty_pic}
		\mathcal{A}\left(\mathcal{C}_a,\mathcal{C}_b\right)= [\mathcal{S}_{\mathcal{C}_{a} \mid \mathcal{C}_{a} \cup \mathcal{C}_{b}}-\mathcal{S}_{\mathcal{C}_{a}}]  + [\mathcal{S}_{\mathcal{C}_{b} \mid \mathcal{C}_{a} \cup \mathcal{C}_{b}}-\mathcal{S}_{\mathcal{C}_{b}} ]
	\end{equation}
	
	where, terms in each square bracket denote the incremental path integral of $\mathcal{C}_{a}$ and $\mathcal{C}_{b}$ respectively.
	$\mathcal{S}_{\mathcal{C}_a}$ is the path integral of the cluster $\mathcal{C}_a$ defined as the weighted sum of transition probabilities of all possible paths from any vertex $i$ to any other vertex $j$. Here $i, j$ vertices belong to the cluster $\mathcal{C}_a$ and all the vertices along the path also belong to cluster $\mathcal{C}_a$. 
	$\mathcal{S}_{\mathcal{C}_{a} \mid \mathcal{C}_{a} \cup \mathcal{C}_{b}}$ is the conditional path integral defined as the path integral of all paths in $\mathcal{C}_{a} \cup \mathcal{C}_{b}$ such that the paths start and end with vertices belonging to  $\mathcal{C}_{a}$. The transition probability matrix $\boldsymbol{P}$ is obtained by normalizing each row of $\bm{W}$ by its sum. 
	% The clusters to be merged at each step are determined using, 
	% \begin{equation}
	%     \left\{\mathcal{C}_{a}, \mathcal{C}_{b}\right\}=\underset{C_{i}, C_{j} \in \mathcal{\textbf{C}}, i \neq j}{\operatorname{argmax}} \mathcal{A}\left(\mathcal{C}_{i}, \mathcal{C}_{j}\right)
	%     \label{eq:ahc_cluster_affinity}
	% \end{equation}
	
	Similar to AHC, we perform merging of two clusters at each time step based on the maximum affinity across all pairs till we reach the required number of speakers. Further details about the algorithm and implementation can be found in \cite{singh2021pic}.

	\subsubsection{VB resegmentation and Overlap detection (VB-overlap)}\label{sec:vb-overlap}
	For further refinement of segment boundaries, we apply VB-HMM resegmentation \cite{diez2018speaker} with posterior scaling \cite{singh2019leap} as described in the baseline \cite{ryant2020dihard}.
	% 9053096
	For overlap detection, we use the overlap detection module available in the pyannote toolkit \cite{Bredin2020}. The neural overlap detection model PyanNet architecture is described in \cite{Bullock2020}; it consists of SincNet filter layers followed by recurrent and fully-connected layers. We use the pre-trained network, trained on the DIHARD-I dataset, to compute the frame-level overlap scores. The overlapping speaker segments identified by the detector are then used to refine the VB-HMM re-segmentation output, similar to the approach described in \cite{Bullock2020}.
	
	\subsection{Narrowband End-to-End system (NES) }\label{sec:nes}
	The architecture of the model is similar to the self-attentive end-to-end neural speaker diarization (SA-EEND) \cite{shinji2019ASRU} combined with encoder-decoder based attractors (EDA) \cite{shinji2020Interspeech}. The model uses $4$ stacked Transformer encoder blocks; each encoder consists of $256$ attention units with $4$ attention heads.
	
	We use $23$-dimensional log-Mel filterbank features, extracted every $10$~ms using a frame length of $25$~ms, with an added context of $\pm 7$ frames. The resulting $345$ dimensional vectors are sub-sampled by a factor of $10$ and used as input to the SA-EEND+EDA model.
	
	\textbf{Training}: To train the narrowband end-to-end system, we simulated $100k$ two-speaker mixtures from Switchboard-2 (Phase I, II, III), Switchboard Cellular (Part 1, Part 2) and NIST SRE datasets 2004-08, using the algorithm proposed in \cite{Fujita2019Interspeech}. The model is trained on the simulated mixtures for $100$ epochs using utterance-level permutation-invariant training (PIT) \cite{PIT_loss} criterion. This is followed by model adaptation on two-speaker recordings from CALLHOME dataset.
	
	\textbf{Evaluation}: For the narrowband speech recordings, we fix the number of attractors to be generated to 2 and obtain the frame-wise posteriors. A threshold is applied to the posteriors to detect the presence of the speakers. We assign the speaker with the maximum posterior in the frames where model does not predict any speaker. The silence frames are then removed based on the SAD. To avoid abrupt speaker change and to make the model more memory efficient, we subsample the frame-level input features with different factors. For SA-EEND V1 and V2, we employ a subsampling factor of $10$ and $5$ respectively.
	% involves subsampling by $10$, whereas SA-EEND V2 involves subsampling by $5$.
	\begin{table}
		\begin{center}
			\caption{DER(JER) performance for wideband system configurations using ETDNN x-vector model and for narrowband using SA-EEND-EDA model indicating the improvements from the proposed approaches for Track-1. $^*$ indicates baseline with oracle number of speakers.}\label{tab:wideband}
			\vspace{-0.2cm}
			% \resizebox{\columnwidth}{!}{%
			\centering
			\begin{tabular}{@{}ccc@{}}
				\toprule
				\textbf{Wideband System config.}                                                                                             & \textbf{Dev DER(JER)}                \\
				\midrule
				
				PLDA+AHC (S1)         & 20.09 (43.86)                        \\
				PLDA + PIC (S2)      &\textbf{18.91 (42.06)}                        \\
				Cosine+ PIC      &{19.78} (43.61)                        \\
				\midrule
				\midrule
				S1+VB-overlap     &    17.70 (42.93)\\
				S2+VB-overlap         &  \textbf{17.01 (41.77)}  \\
				\midrule
				\midrule
				\textbf{Narrowband System config.}                                                                                             & \textbf{Dev DER(JER)}                \\
				\midrule
				Baseline w Oracle$^*$       &16.03 (20.21)\\
				SA-EEND V1        & \textbf{9.84 (12.00)}\\
				SA-EEND V2          &\textbf{9.34 (11.19)}\\
				
				\bottomrule
				\vspace{-2em}
			\end{tabular}%
			% }
			\vspace{-0.4cm}
		\end{center}
	\end{table}
	\section{Data Resources}\label{sec:data_resources}
	DIHARD III dataset \cite{ryant2020dihard} comprises of 0.5-10 minutes of recordings drawn from 11 different domains ranging from audiobooks to web videos. It consists of variations in number of speakers (1-10), overlapping speech (0-80\%) and diverse speaker turn behavior, making the task of speaker diarization more challenging. The dataset is divided into dev and eval set for training and evaluation respectively, containing about 60 hours of audio. Each set is further divided into full and core for evaluation. The full set considers entire set for evaluation, whereas the core set excludes some files from domains with more hours of audio so that all domains contribute equally to the overall performance. Training datasets are listed here.
	\subsection{Wideband system datasets}
	\begin{itemize}
		\item VoxCeleb1 \cite{nagrani2017voxceleb} and VoxCeleb2 \cite{Chung2018}: The two datasets respectively contains over 0.1 and 1 million utterances from thousands of celebrities, extracted from videos uploaded to YouTube. Both the dataset are fairly gender balanced with 55\% and 61\% of the speakers belonging to the male gender respectively. The video recordings included in the datasets are recorded in a large number of challenging visual and auditory environments.
		
		% \item VoxCeleb1\footnote{\url{http://www.robots.ox.ac.uk/~vgg/data/voxceleb}}\cite{nagrani2017voxceleb}: It contains over $100k$ utterances from 1,251 celebrities, extracted from videos uploaded to YouTube. The dataset is gender balanced, with 55\% of the speakers being male. The speakers span a wide range of ethnic backgrounds, accents, professions and ages.
		% \item VoxCeleb2\footnote{\url{http://www.robots.ox.ac.uk/˜vgg/data/voxceleb2}}\cite{Chung2018}: It  contains over $1$ million utterances from over 6,000 celebrities, extracted from videos uploaded to YouTube. The dataset is fairly gender balanced, with 61\% of the speakers belonging to male gender. The video recordings included in the dataset are recorded in a large number of challenging visual and auditory environments.
	\end{itemize}
	\subsection{Narrowband system datasets}
	
	Telephone recordings: We use Switchboard-2 (Phase I, II, III), Switchboard Cellular (Part 1, Part 2) and
	NIST SRE datasets 2004-2008.\\
	CALLHOME (CH) : It is a collection of multi-lingual telephone call recordings sampled at 8 kHz, containing 500 recordings. The duration of each recording ranges from 2-5 minutes. The number of speakers in each recording varies from 2 to 7, and majority of the recordings have two speakers. The CH dataset
	is divided equally into two different sets, CH1 and CH2,
	with similar distribution of number of speakers.
	
	\begin{figure}[t!]
		\centering
		\includegraphics[trim={0cm 0cm 0cm 0cm},clip,width=\linewidth,height=4.72cm]{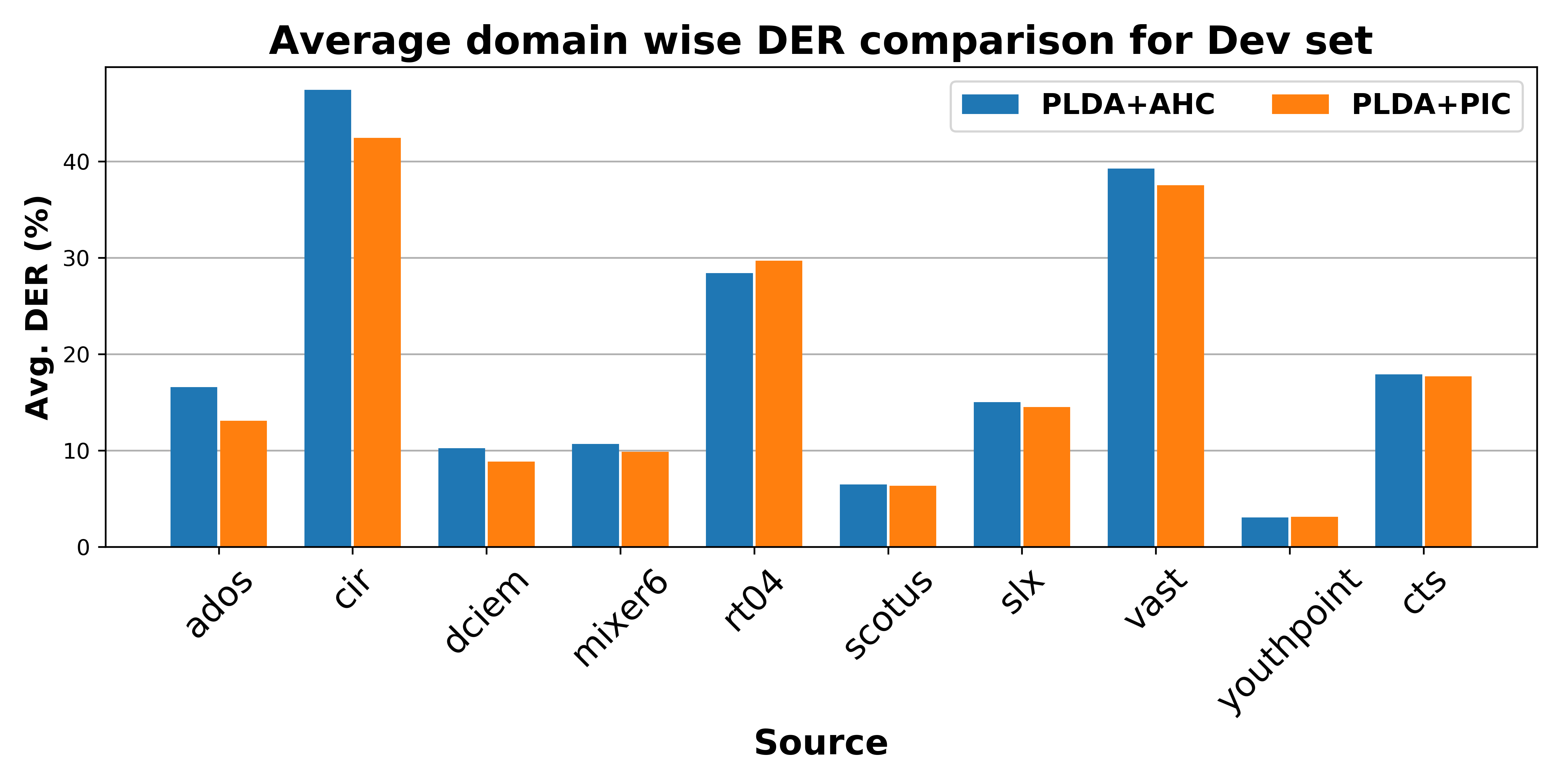}
		\caption{Performance comparison of AHC and PIC algorithms across domains in DIHARD dev set.}
		\label{fig:domain}
	\end{figure}
	\section{Experiments \& Results}
	The system performance is evaluated using the diarization error rate (DER) and Jaccard error rate (JER) measures, computed using the \textit{dscore} toolkit \cite{ryant2020third}.
	% It calculates the sum of missed speech, false alarm and speaker error rate.
	% Additionally,  is also used as a secondary metric.More details can be found in DIHARD III Evaluation Plan \cite{ryant2020third}. 
	
	We have applied different strategies for wideband and narrowband subset of DIHARD dev set. For wideband, we experiment with different scoring and clustering techniques. Table \ref{tab:wideband} shows DER(JER) performance of wideband system using ETDNN x-vectors. The PLDA+AHC is the baseline approach \cite{ryant2020dihard}. As discussed in Section \ref{sec:pic}, we have implemented PLDA and cosine with PIC. From Table \ref{tab:wideband}, we observe that PLDA with PIC algorithm gives 1\% absolute improvement compared to PLDA with AHC algorithm. The stopping criteria of PIC is based on number of speakers predicted by PLDA+AHC threshold obtained after fine tuning on dev set. The comparison of PIC and AHC is elaborated in Figure \ref{fig:domain}. It shows domain wise average DER for PLDA-AHC and PLDA-PIC using the ETDNN embedding extractor. The plot shows improvements for PIC algorithm over AHC in 7 out of 9 domains in wideband. Further refinement using VB-HMM and overlap detection gave an improvement of $3\%$ absolute improvement over the baseline system. The miss-rate and false-alarm of the overlap-detection module on DIHARD dev set are 17.1\% and 5.0 \% respectively and on DIHARD eval set are 56.68\% and 2.91\% respectively.
	
	%  For evaluation, we subsample the frame-level features with different factors to avoid abrupt speaker change and to make it more memory efficient. Table \ref{tab:wideband} also shows results on narrowband recordings from CTS domain. SA-EEND V1 involves subsampling by 10, whereas SA-EEND V2 involves subsampling by 5. 
	For narrowband system, we use SA-EEND system with attractor. Table \ref{tab:wideband} also shows results on narrowband recordings from CTS domain. We see a significant improvement of 40\% (relative) for SA-EEND model over the baseline. As we increase the resolution by reducing subsampling factor, we only observe a marginal improvement.
	
	Tables \ref{tab:track1} and \ref{tab:track2} show results of our systems and the baseline for Track-1 and Track-2 respectively. We have combined WPS with ETDNN/FTDNN for wideband domain files and NES for narrowband (CTS domain) files. The classification accuracy of wideband-narrowband classifier for eval set is 99.67\%. In Track-1, we obtain 24\% and 15\% relative improvements in DER for full and core sets respectively using ETDNN model over the baseline. A similar performance gain is obtained using the FTDNN model also. In Track-2, we obtain around 18\% and 11\% relative improvements in DER for both ETDNN and FTDNN model for full and core sets over the baseline. 
	% We also explored weighted combination of PLDA scores from ETDNN and FTDNN models for wideband PIC system (WPS) which only gave marginal improvement. 
	\begin{table}
		\centering
		\caption{Performance of individual systems  on Track-1 .} \label{tab:track1}
		\vspace{-0.2cm}
		\resizebox{\columnwidth}{!}{%
			\begin{tabular}{@{}cccc@{}}
				\toprule
				\textbf{Individual System}          & \textbf{Set}         & \textbf{Dev DER(JER)} & \textbf{Eval DER(JER)} \\ 
				\midrule
				Baseline\cite{ryant2020dihard}  &full & 19.10 (41.10)        & 19.68 (44.32)         \\
				&core & 19.97 (45.52)        & 21.35 (48.89)         \\
				WPS (ETDNN)+NES  &full &  14.45 (\textbf{37.09})        &  14.93 (37.09)         \\
				&core  &16.43 (\textbf{42.45})        &18.2 (43.28)         \\
				WPS (FTDNN)+NES  &full & \textbf{14.34} (37.31)       &     \textbf{14.88 (36.73)}    \\
				&core  &  \textbf{16.26} (42.75)       &      \textbf{18.07 (42.82)}    \\
				
				% \textbf{Fused system[\ref{sec:fusion}}] &\textbf{Set} &\textbf{Dev DER(JER)} & \textbf{Eval DER(JER)} \\ 
				% \midrule
				% WPS(ETDNN + FTDNN) + NES &full &  () &  () \\
				%                         &core & () &  () \\
				
				\bottomrule
			\end{tabular}%
		}
		\vspace{-0.2cm}
	\end{table}
	
	\begin{table}
		\centering
		\caption{Performance of individual systems on Track-2 .} \label{tab:track2}
		\vspace{-0.2cm}
		\resizebox{\columnwidth}{!}{%
			\begin{tabular}{@{}cccc@{}}
				\toprule
				\textbf{Individual System}          & \textbf{Set}         & \textbf{Dev DER (JER)} & \textbf{Eval DER (JER)} \\ 
				\midrule
				Baseline\cite{ryant2020dihard}  &full & 
				21.35 (42.97)        & 25.76 (47.64)         \\
				&core & 22.31 (47.28)        & 28.31 (52.44)         \\
				WPS(ETDNN)+NES  &full &  16.77 (\textbf{37.15})        &  21.04 (39.68)         \\
				&core &  18.64 (\textbf{41.93})        &  24.92 (45.32)         \\
				WPS(FTDNN)+NES  &full &\textbf{16.53} (38.50)          &    \textbf{21.09 (39.54)}       \\
				&core & \textbf{18.34} (43.62)      &     \textbf{24.99 (45.13)}     \\
				% \midrule
				% \midrule
				
				% \textbf{Fused system[\ref{sec:fusion}}] &\textbf{Set} &\textbf{Dev DER(JER)} & \textbf{Eval DER(JER)} \\ 
				% \midrule
				% WPS(ETDNN + FTDNN) + NES &full &  () &  () \\
				%                         &core & \textbf{20.56} (47.43) & \textbf{21.90} (49.93) \\
				
				\bottomrule
			\end{tabular}%
		}
		\vspace{-0.4cm}
	\end{table}
	\section{Post Evaluation Analysis}\label{sec:post_dihard}
	Here, we discuss details and results of our post evaluation analysis to improve performance of wideband system for Track-1.
	\subsection{VBx resegmentation}
	We perform hidden Markov model (HMM) based clustering on x-vectors using variational Bayes (VB) approach with VBx clustering \cite{9053982}. It is a simplified version of VB-HMM model in which the state distributions are Gaussian and the observations are the x-vectors of the recording. Each state represents a speaker and transition probabilities across states represents speaker turn. The loop probability $Ploop=0.8$ is the probabilty to remain in the same speaker state. The parameters of each Gaussian are obtained from a pre-trained PLDA model. As per \cite{9053982}, we train two PLDA models using VoxCeleb dataset and DIHARD dev set using x-vectors from $3$~s segments. The pre-processing steps include whitening using $1.5$~s DIHARD dev x-vectors with $0.25$~s shift followed by length normalization. The x-vectors are then projected
	from $512$D to $220$D using Linear Discriminant Analysis (LDA). We perform interpolation of both PLDAs regulated by $\alpha=0.5$. The AHC is used to initialize the HMM resegmentation. We have experimented with ETDNN model based on MFCC and mel filterbank features. The best results are obtained using mel filterbank features.
	
	\subsection{End-to-End Wideband system}
	We have explored the use of end-to-end system for other domains in DIHARD III excluding CTS. The architecture of the model is same as the one discussed in Section \ref{sec:nes}.\\
	\textbf{Training:} We simulated $400k$ mixtures for 1-4 speakers from VoxCeleb1 and 2 and Librispeech \cite{panayotov2015librispeech} datasets. We trained the model on $2$ speaker mixtures for $12$ epochs followed by fine tuning on $1$-,$2$-,$3$-,$4$- speaker mixtures for $4$ epochs. Finally, the model is adapted on DIHARD III dev set.\\
	\textbf{Evaluation:} We followed the same procedure for evaluation as described in section \ref{sec:nes}, except that instead of fixing the number of attractors to be generated to $2$, we let the model infer the required number based on the attractor existence probabilities.
	
	% \subsection {SSC-PLDA-PIC}

	\subsection{Results}\label{sec:expt_results}

	% \begin{table}
	% \begin{center}
	% \caption{DER(JER) performance for wideband system configurations using Fbank-ETDNN x-vector model and for narrowband using SA-EEND model indicating the improvements from the proposed approaches for track1. $^*$ indicates baseline with oracle number of speakers.}\label{tab:wideband}
	% \vspace{-0.2cm}
	% % \resizebox{\columnwidth}{!}{%
	% \centering
	% \begin{tabular}{@{}ccc@{}}
	% \toprule
	% \textbf{Wideband System config.}                                                                                             & \textbf{Dev DER(JER)}                \\
	% \midrule
	
	% PLDA+AHC+VBx (S1)         & 16.22 (37.67)                        \\
	% PLDA + PIC (S2)      &18.00 (41.92)                        \\
	% \midrule
	% \midrule
	%  S1+overlap     &     14.50 (37.25)\\
	%  S2+overlap         &  15.20 (38.38)  \\
	
	%  \bottomrule
	% \vspace{-2em}
	% \end{tabular}%
	% % }
	% \vspace{-0.4cm}
	% \end{center}
	% \end{table}
	\begin{table}
		\centering
		\caption{Post DIHARD analysis: The DER (JER) performance of individual systems  on Track-1 .} \label{tab:track1_post_dihard}
		\vspace{-0.2cm}
		\resizebox{\columnwidth}{!}{%
			\begin{tabular}{@{}cccc@{}}
				\toprule
				\textbf{Individual System}          & \textbf{Set}         & \textbf{Dev DER(JER)} & \textbf{Eval DER(JER)} \\ 
				\midrule
				LEAP Eval Sub.  &full & 14.34 (37.31)        & 14.88 (36.73)         \\
				&core & 16.26 (42.75)       & 18.07 (42.82)         \\
				
				WB VBx-ovp + NES &full &\textbf{12.87} (\textbf{33.11}) & \textbf{14.53 (35.62)} \\
				&core & \textbf{14.26} (\textbf{37.64}) & \textbf{17.45 (40.90)} \\
				WPS (Mel)+VBx + NES &full &13.22 (34.05) &  15.04 (38.17) \\
				&core & 14.71 (38.68) & 18.20 (44.08) \\
				WB EEND + NES & full & 19.19 (47.17) & 18.66 (44.78)  \\
				&core & 23.43 (55.08)  & 23.55 (52.59) \\
				\bottomrule
			\end{tabular}%
		}
		\vspace{-0.6cm}
	\end{table}
	Table \ref{tab:track1_post_dihard} shows our contributions post DIHARD evaluation. VBx system combined with overlap detection gave best results for wideband on dev set. We combined wideband VBx with Narrowband end-to-end system (NES) to get best performance. We get 26\% and 18\% relative improvements over baseline in full and core eval sets respectively. The third row in the Table \ref{tab:track1_post_dihard} shows results of WPS system with ETDNN trained with mel filterbank and interpolated PLDA where we obtain the number of speakers from VBx system. Owing to large diversity in the domains and the training complexity, the wideband end-to-end system (last row of Table \ref{tab:track1_post_dihard}) generated suboptimal results.
	\section{Conclusions}\label{sec:conclusions}
	We have presented the detailed description of the LEAP submission to DIHARD-III diarization challenge. The key novel components of the system include, (i) a band classifier based approach to separately process conversational telephone data and the other wide-band domains, (ii) graph based clustering approach to diarization using path integral clustering and (iii) incorporation of overlap detector with the final re-segmentation step to improve the diarization performance. The paper also includes our analysis further to the conclusion of the challenge where we observe additional performance improvements with the use of re-segmentation based on the x-vector embeddings instead of frame-level features. 
	\section{Acknowledgement}
	The authors would like to acknowledge the technical discussions with Abhishek Anand, Rohit Singh and Dr. Michael Free of BT Research that helped in shaping the paper. The authors would like to thank Hitachi-JHU DIHARD team for providing pre-trained ETDNN model for the post DIHARD analysis.
	\bibliographystyle{IEEEtran}
	
	\bibliography{template}

\end{document}